\documentclass[aps,pra,twocolumn,notitlepage,superscriptaddress,showpacs,10pt]{revtex4-1} 
\usepackage{amsmath}    
\usepackage{mathtools} 
\usepackage{amsfonts}
\usepackage{bm}
\usepackage{amssymb}
\usepackage{appendix}
\usepackage{graphicx}   
\usepackage{xcolor}      
\usepackage{subfigure}  
\usepackage{comment}
\usepackage[separate-uncertainty=true]{siunitx}
\usepackage{cancel}

\usepackage{soul}
\usepackage[normalem]{ulem}

\mathchardef\mhyphen="2D
\begin{document}

\title{Optical control of topological end states via soliton formation in a 1D lattice}
\author{Christina J{\" o}rg}
\email[]{cjoerg@rptu.de}
\affiliation{Department of Physics, The Pennsylvania State University, University Park, Pennsylvania 16802, USA}
\affiliation{Physics Department and Research Center OPTIMAS, University of Kaiserslautern-Landau, Kaiserslautern D-67663, Germany}
\author{Marius J{\" u}rgensen}
\affiliation{Department of Physics, The Pennsylvania State University, University Park, Pennsylvania 16802, USA}
\author{Sebabrata Mukherjee}
 \affiliation{Department of Physics, The Pennsylvania State University, University Park, Pennsylvania 16802, USA}
\affiliation{Department of Physics, Indian Institute of Science, Bangalore 560012, India}

\author{Mikael C. Rechtsman}
\affiliation{Department of Physics, The Pennsylvania State University, University Park, Pennsylvania 16802, USA}

\date{\today}

\begin{abstract}

Solitons are self-consistent solutions of the nonlinear Schr\"odinger equation that maintain their shape during propagation.
Here we show, using a pump-probe technique, that soliton formation can be used to optically induce and control a linear topological end state in the bulk of a Su-Schrieffer-Heeger lattice, using evanescently-coupled waveguide arrays. 
Specifically, we observe an abrupt nonlinearly-induced transition above a certain power threshold due to an inversion symmetry-breaking nonlinear bifurcation. 
Our results demonstrate all-optical active control of topological states.
\end{abstract}

\maketitle

In topological systems, bulk invariants ensure robustness against local defects and disorder via protected states residing at the physical boundaries \cite{lu_topological_2014,ozawa_topological_2019,SegevBandres+2021+425+434}, implying the possibility of optical devices more robust to fabrication-induced disorder and damage \cite{bandres_topological_2018,PhysRevLett.120.113901,PhysRevLett.122.153904}. 
While topological effects in linear systems have been well studied, 
nonlinear topological systems have only been explored very recently
\cite{PhysRevLett.111.243905,PhysRevLett.117.143901,Rachel_2018,kruk_nonlinear_2019,smirnova_nonlinear_2020,xia2020nontrivial,maczewsky2020nonlinearity,SebaFloquetSoliton, PhysRevB.102.115411, xia2021nonlinear,kirsch2021nonlinear, SolitonEdgeStates,IntegerThouless, pernet_gap_2022,PhysRevLett.128.093901,ABLOWITZ2022133440,mukherjee2022period,FractionalThouless,bai2024arbitrarily,szameit_discrete_2024}, 
and a theoretical framework is just starting to emerge~\cite{PhysRevB.102.115411,sone2023nonlinearityinduced,PhysRevB.108.195142}. 
Perhaps most understanding has been gained for nonlinear topological systems based on spatial solitons, which are self-consistent solutions of the nonlinear Schrödinger equation that maintain their spatial shape during propagation. At high optical power the nonlinear detuning of the refractive index (e.g. Kerr effect) balances the diffraction determined by the hopping between neighboring sites. Using waveguide arrays, spatial solitons \cite{Christodoulides:88,PhysRevLett.68.923,PhysRevLett.81.3383,PhysRevLett.90.023902} have been observed in the bulk of anomalous Floquet topological insulators \cite{SebaFloquetSoliton,mukherjee2022period}, propagating along the edge \cite{SolitonEdgeStates}, or being pumped by integer and fractionally quantized values in photonic Thouless pumps \cite{IntegerThouless,FractionalThouless}.

Here, we use optical soliton formation to nonlinearly induce and probe a topological end state (that is, a 0D topological edge state) at any position in the bulk of a Su-Schrieffer-Heeger (SSH) lattice. 
In contrast to previous works that examined the role of nonlinearity on linear topological end states \cite{PhysRevB.93.155112,PhysRevLett.121.163901,Guo:20,PhysRevE.104.054206,PhysRevB.105.035410,arkhipova2022observation}, in our system the soliton optically induces a topological state in the bulk of the sample, not at a physical termination of the lattice. 
To this end, we use an all optical pump-probe setup with zero time delay and orthogonally polarized beams.
We use the high-power pump beam to generate a soliton in the bulk of evanescently coupled waveguides that acts as a hard wall, while the low-power probe beam probes the end state that is induced next to the wall. The soliton therefore acts as an all-optical switch to turn on and off the topological (linear) end state. We theoretically show that the end state formation occurs above a certain threshold power due to a spontaneous inversion breaking nonlinear bifurcation. We complete our analysis by connecting the end states with previously (see Ref.~\cite{eilbeck_discrete_1985}) known analytical solutions for dimers.


\begin{figure}[t]
    \centering
    \includegraphics[width=\linewidth]{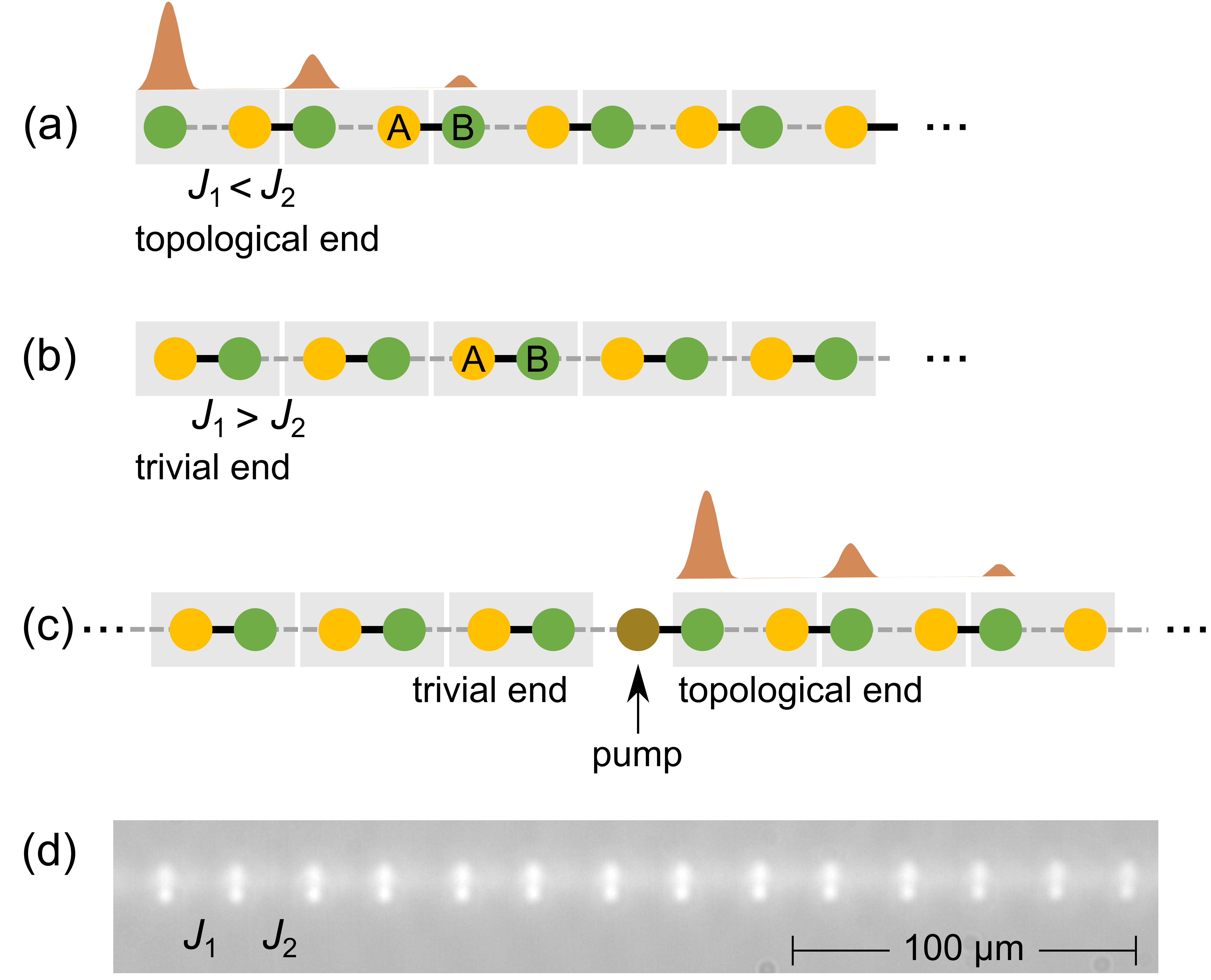}
    \caption{ \small{
        \textbf{Schematic of the SSH lattice.}
        (a) Topological termination for first hopping $J_1<J_2$. (b) Trivial termination for first hopping $J_1>J_2$. Boxes indicate the unit cells. (c) Nonlinear chain (in the infinite power limit) where the refractive index of the pumped site is being detuned, creating a topological ending to the right of the pumped site, while there is a trivial ending to the left of it. The sublattices are labeled A and B for easier identification, but do not have any (linear) onsite potential difference. (d) White-light image of the output facet of the SSH waveguide array.
      \label{fig:0}}}
\end{figure}

We demonstrate our results using a SSH model -- a dimer lattice, consisting of an A and B sublattice, with intra-cell hopping $J_1$ and inter-cell hopping $J_2$ ($J_1$,$J_2\geq 0$), depicted in Fig.~\ref{fig:0}. For a finite chain, topological end states are present and exponentially localized provided the final hopping of the lattice is the weaker of $J_1$ and $J_2$. 
In coupled waveguide systems and at high optical power, nonlinearity emerges as an intensity-dependent modification of the waveguide's refractive index (and therefore of the on-site potential, in contrast to nonlinear couplings \cite{Zhou_2017,PhysRevB.93.155112,PhysRevB.100.014302}) due to the Kerr effect. The dynamics in our system for the pump and probe beam are then described by the following coupled equations:
\begin{align}
    \mathrm{i} \partial_z \Psi_n^{\mathrm{pu}} = \sum_m H^{\mathrm{lin}}_{n,m} \Psi_m^{\mathrm{pu}} - g|\Psi_n^{\mathrm{pu}}|^2\Psi_n^{\mathrm{pu}}
    \label{eq:1}
\end{align}
\begin{align}   
    \mathrm{i} \partial_z \Psi_n^{\mathrm{pr}} = \sum_m  H^{\mathrm{lin}}_{n,m} \Psi_m^{\mathrm{pr}}- g|\Psi_n^{\mathrm{pu}}|^2  \Psi_n^{\mathrm{pr}},
    \label{eq:2}
\end{align}
where $H^{\mathrm{lin}}$ is the tight-binding Hamiltonian of the linear SSH system, containing the hopping amplitudes $J_1$ and $J_2$, and $\Psi_n^{\text{pu}}(z)$ and $\Psi_n^{\text{pr}}(z)$ are proportional to the electric field amplitudes of the pump and probe beam at site $n$ and propagation distance $z$. 
The total normalized power ${\cal P}_{\text{pu (pr)}}= \sum_n|\Psi_n^{\text{pu (pr)}}|^2$ in the pump (probe) polarisation is conserved. 
The nonlinear parameter $g$ is defined as $g = 2 \pi n_2 P/( \lambda A_\mathrm{eff})$, where $n_2$ is the nonlinear refractive index, $P$ is the total optical power of the beam, and $A_\mathrm{eff}$ is the effective area of the waveguide mode, such that $gP/J_1$ is dimensionless.
Due to the vastly different powers of the beams with orthogonal polarizations we neglect the Kerr effect as induced by the probe beam, and only consider that of the pump. The pump's dynamics (Eq.~\ref{eq:1}) is described by the discrete nonlinear Schr\"odinger equation (corresponding to the Gross-Pitaevskii equation for the description of Bose-Einstein condensates and superfluids). 
The probe beam's dynamics (Eq.~\ref{eq:2}) is linear but dependent on the pump's induced nonlinear potential.  Therefore, we first focus on the formation of the soliton in the pump polarization at high optical power; later we will show its impact on the low-power orthogonally polarized probe beam. 


\begin{figure*}
    \centering
    \includegraphics[width=\linewidth]{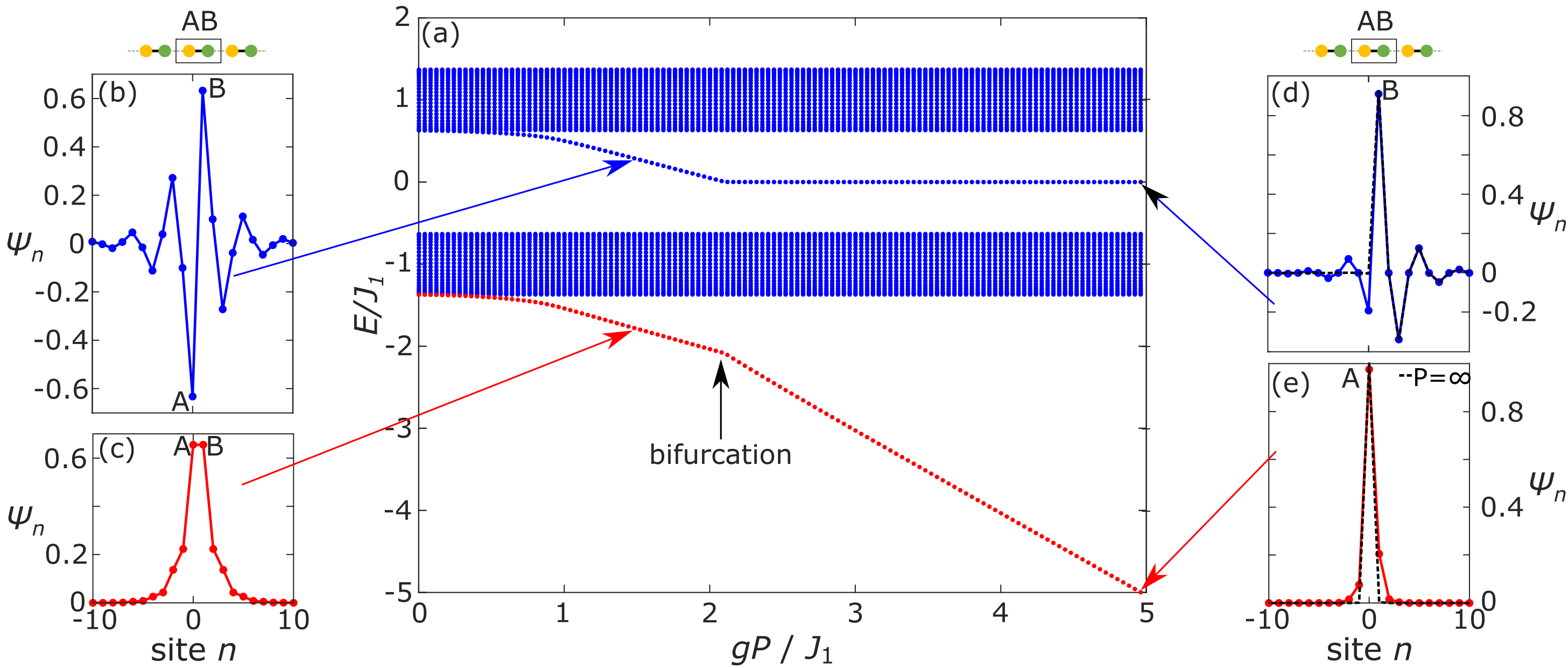}
    \caption{ \small{
        \textbf{Spectrum obtained via the self-consistency method}
         using 100 sites, $J_2/J_1=0.37$ and open boundary conditions with topologically trivial terminations. (a) The nonlinear eigenvalue of the soliton is plotted in red, and the eigenvalues of Eq.~\ref{eq:2} under the potential of the obtained pump soliton wavefunction are plotted in blue. Clearly visible is the soliton bifurcation point at power $g P/J_1 \approx 2$, where the soliton eigenvalue changes its slope for increasing power, and the eigenvalue of the  in-gap state goes to zero.  The wavefunctions of the soliton (red, (c) and (e)) and the in-gap state (blue, (b) and (d)) are shown for low power (left) and high power (right). For low power, both states have equal support on the A and B sublattices. For $g P/J_1 \gtrapprox 2$ the soliton symmetry is broken and the soliton has greater support on the input site (A in this case). The in-gap state then turns into a topological end state with support mostly on the B sublattice. The dashed black lines in (d) and (e) indicate the limit of infinite input power. 
      \label{fig:1a}}}
\end{figure*}


We solve Eq.~\ref{eq:1} using the self-consistency method \cite{PhysRevLett.79.4990} to find localized soliton solutions. Figure~\ref{fig:1a}(a) shows the calculated spectrum. The soliton energy eigenvalue is plotted in red, and blue dots show the eigenvalues of Eq.~\ref{eq:2} under the potential defined by the pump soliton wavefunction.  In other words, this calculation contains both the soliton itself, as well as the linear solutions that appear as a result of the presence of the soliton. 
We simultaneously observe two effects: For power $gP\approx2J_1$ the soliton eigenvalue abruptly changes its slope as a function of power, and the in-gap state's eigenvalue also abruptly approaches mid-gap (zero eigenvalue). The corresponding eigenstates (Fig.~\ref{fig:1a}(b),(c),(d),(e)) reveal that for low power, the soliton wavefunction (red) has equal support on both sublattices, as the soliton wavefunction is inversion symmetric (see also Refs.~\cite{arkhipova2022observation,PhysRevE.65.056607,PhysRevA.79.065801,https://doi.org/10.1002/lpor.201900223}), but at the transition, the soliton inversion symmetry is broken and gains more support on a single site for increasing power: The soliton undergoes a spontaneous inversion symmetry breaking nonlinear bifurcation (see Supplementary Information section III). 

Furthermore, the in-gap state has equal support on both sublattices below the bifurcation threshold, but its support on the A sublattice becomes very small by comparison above the threshold. This is highly reminiscent of a topological end state in the SSH lattice: in systems that respect chiral symmetry, localized zero modes are supported on only one sublattice. Note that, in the limit of infinite power the soliton wavefunction only has support on a single site and the wavefunction of the linear in-gap state exactly matches that of a topological end state (dashed black lines in Fig.~\ref{fig:1a}(d),(e)). We understand this behavior as follows: in this limit, the pump light detunes the input waveguide's refractive index sufficiently strongly that it effectively acts as a wall, cutting the SSH lattice into two subchains (see Fig.~\ref{fig:0}(c)), with a trivial termination on the left and a topologically non-trivial termination on the right.

The fully dimerized limit of the lattice (i.e., when $J_2=0$) allows us to understand the sharp transition around power $g P \approx 2J_1$. In this regime, we may consider the ``lattice'' as being simply composed of individual dimers (i.e., pairs coupled sites that are uncoupled to others). For a single dimer, for $g P<2J_1$, one symmetric and one anti-symmetric nonlinear mode exists. However, for $g P>2J_1$ the symmetric mode becomes unstable through a symmetry-breaking nonlinear bifurcation (see also Supplementary Information section III for a linear stability analysis) and a new stable nonlinear mode emerges that has dominant support on one of the two dimer sites \cite{eilbeck_discrete_1985}. The linear eigenstate of Eq.~\ref{eq:2} that includes the potential induced by the soliton then has an eigenvalue of exactly zero (see Supplementary Information section III). When moving away from the dimerized limit, these become the localized soliton mode and the mid-gap  end state, respectively. As we show in Supplementary Information Fig. S5, as $J_2$ is increased from zero, the transition remains, but shifts to slightly higher power. In our experiment, we use a pump beam to induce the soliton, which then defines the trapping potential for the linear end state, which we probe in the orthogonal polarization.

\section{Experiment \label{Experiment}}

\begin{figure*}[t]
    \centering
    \includegraphics[width=\linewidth]{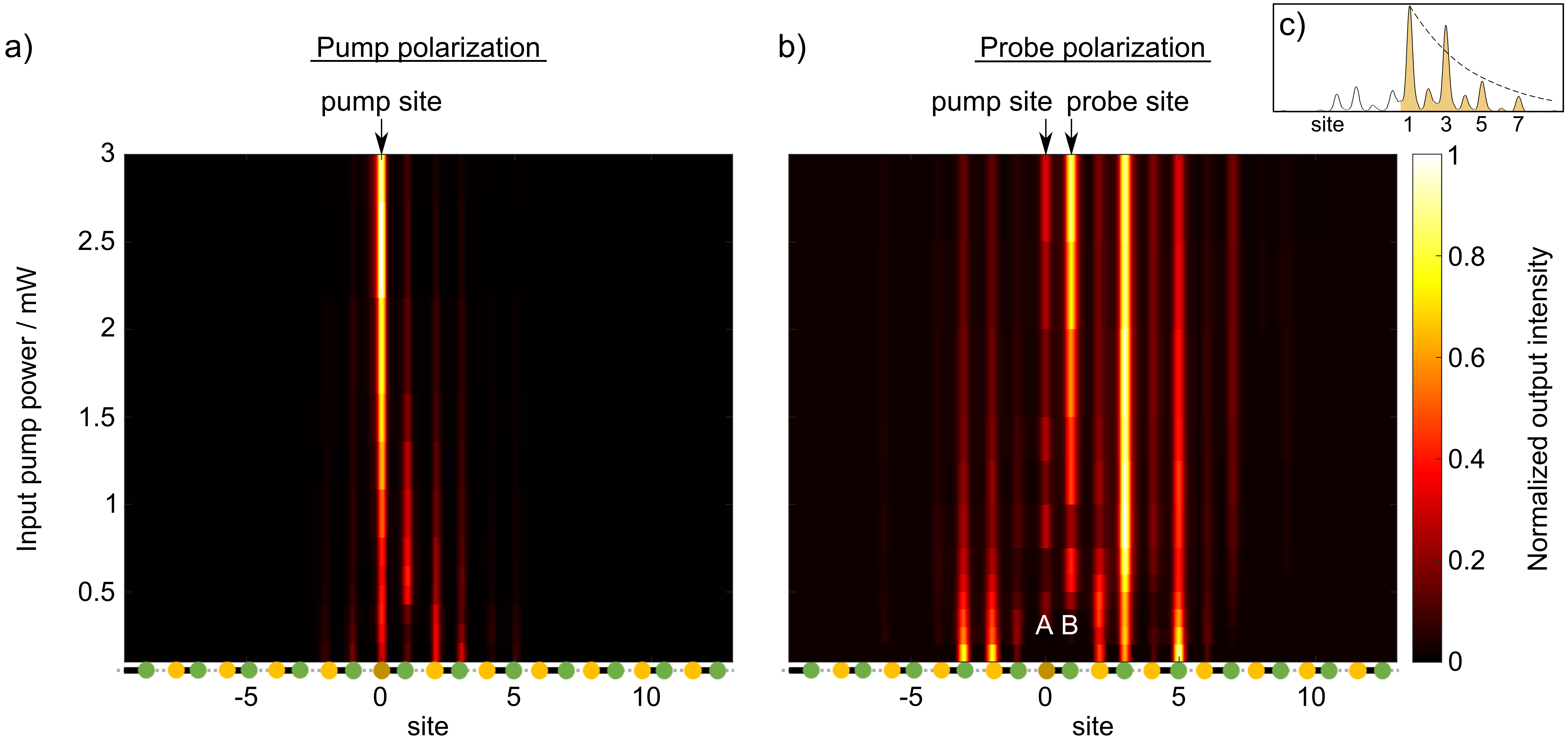}
    \caption{ \small{
        \textbf{Experimental observation of nonlinearly-induced topological end states.}
        (a) Pump intensity at the output facet for increasing pump power. The pump beam is injected into site 0. 
        (b) Probe intensity at the output facet for increasing pump power. The pump beam is injected into site 0, while the probe beam is focused into the neighboring strongly-coupled site (site 1). For low power the probe intensity shows equal support on both sublattices, while for high power we see a stronger support on the B sublattice.
        The inset (c) shows the intensity profile (solid line) for the highest power reached in experiment, along with a fit for exponential decay of the intensity towards the bulk (dashed line).
      \label{fig:3ab}}}
\end{figure*}

\begin{figure}
    \centering
    \includegraphics[width=\linewidth]{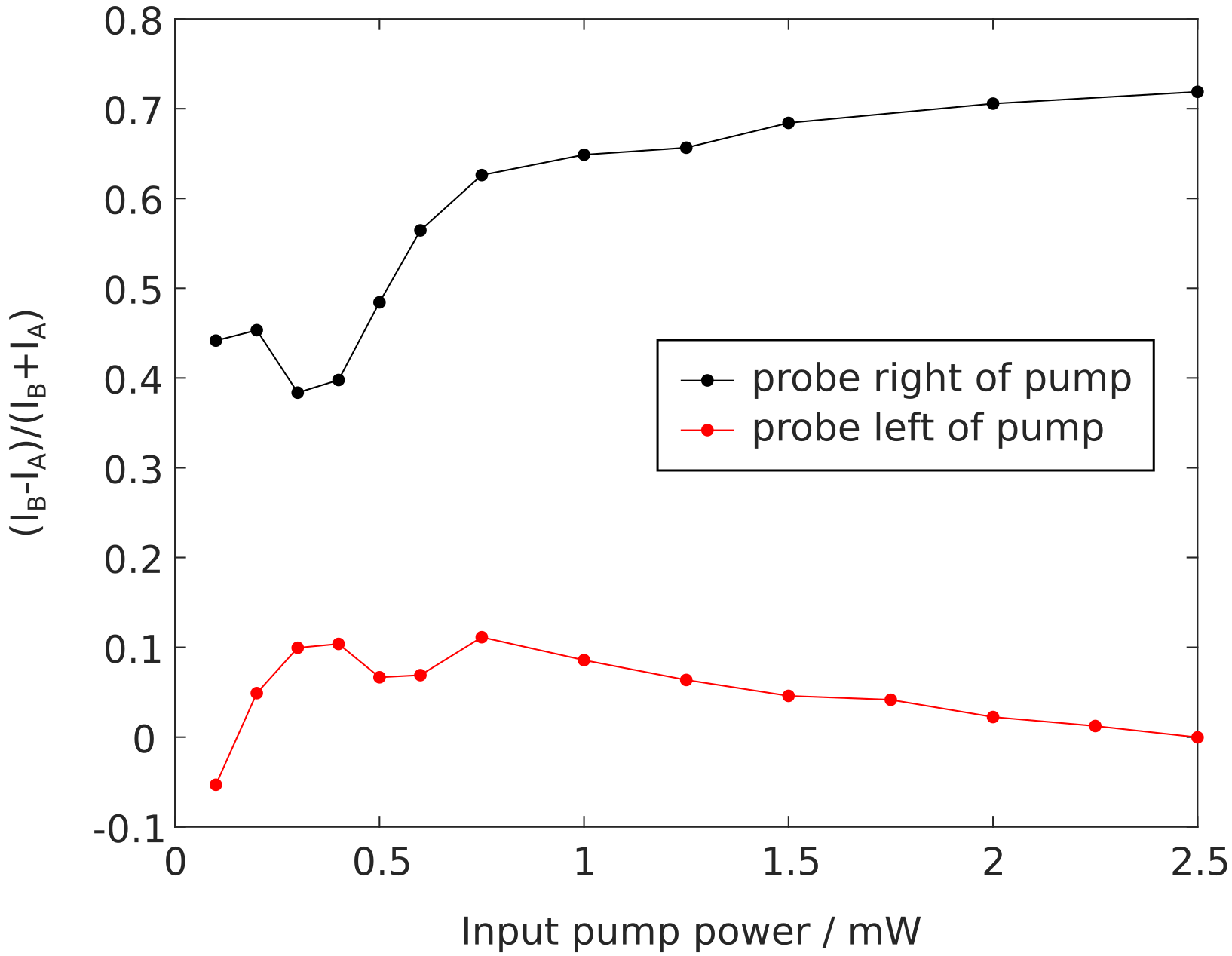}
    \caption{ \small{
        \textbf{Intensity imbalance in the formation of topological end states.}
        Extracted intensity imbalance between A and B sublattice $(I_\mathrm{B}-I_\mathrm{A})/(I_\mathrm{B}+I_\mathrm{A})$. Black dots: Probe beam is injected to the right of the pump (waveguide 1), i.e., the configuration shown in Fig.~\ref{fig:3ab}(a), exciting an induced topological end state. Red dots indicate the imbalance when the probe is injected to the left of the pump (waveguide -1), i.e., not exciting a topological end state. Lines are guides to the eyes. 
      \label{fig:3c}}}
\end{figure}

We experimentally probe the soliton-induced end state in the bulk of a nonlinear SSH chain using single-moded, femtosecond laser written waveguides in Corning Eagle XG borosilicate glass (for more details on the fabrication method see the Supplementary Information section I and Ref.~\cite{szameit_discrete_2010, SebaFloquetSoliton}). To maximize the effective nonlinearity ($g P/J_1$), we separate the waveguides such that the evanescent couplings $J_1$ and $J_2$ are small, while also keeping the coupling constants large enough to see enough transverse dynamics over the maximum propagation length of $\SI{150}{mm}$. The experimental parameters are: $J_{1,\mathrm{pr}}\approx \SI{0.039\pm0.006}{mm^{-1}}$ and $J_{2,\mathrm{pr}}\approx \SI{0.016\pm0.003}{mm^{-1}}$ for the probe beam, and 
$J_{1,\mathrm{pu}}\approx \SI{0.022\pm0.004}{mm^{-1}}$ and $J_{2,\mathrm{pu}}\approx \SI{0.008\pm0.002}{mm^{-1}}$ for the pump beam; the hoppings for the pump and probe beam are different because the polarizations are different. $gP/J_1$ is then scaled to $J_{1,\mathrm{pu}}$ (dropping the index pu), since the pump power determines the bifurcation threshold. In order to reach the necessary degree of nonlinearity, we use high power laser pulses (see Supplementary Information section II). 

The experimental setup is depicted in Supplementary Fig. S1. We use a pump-probe scheme, in which we split the emitted pulses from a single laser source into two orthogonally polarized beams, one with high and the other with low power. We use the high-power beam as the pump beam that excites a soliton, and simultaneously probe the induced end state using a weak-power probe beam in the neighboring waveguide. Both beams are focused into single waveguides using the same lens. Temporal overlap of the pump and probe beam is assured by adjusting the probe beam path with the help of a delay stage. Finally, we image the waveguide intensities at the output facet of the orthogonally polarized pump and probe beam onto separate cameras, using a thin film polarizer. In the experiment, the power of the probe beam is fixed at approximately 0.07 times the pump beam power and therefore is always low enough to ensure that it does not generate any nonlinear effects itself. We estimate the nonlinear length -- the propagation distance for which nonlinear effects are significant -- for the probe beam at its highest power to be $1/g=\SI{173}{mm}$, which is larger than the length of our sample. 

We focus the pump beam into a waveguide in the lattice (indexed by site 0 in Fig.~\ref{fig:3ab}(a)) and the probe beam into the strongly coupled neighboring site, waveguide 1. Fig.~\ref{fig:3ab}(a) shows the output intensity distribution for the pump beam as a function of increasing input power (see also Eq.~\ref{eq:1}). For low input power the intensity is distributed over multiple sites. At high power the wavefunction is strongly peaked on a single site as the symmetry-broken soliton forms and is efficiently excited due to the large mode overlap with a single-site excitation. The soliton acts as the hard wall to split the SSH chain. Note that due to the single-site excitation, we do not efficiently excite the symmetric soliton at low power. 

Figure~\ref{fig:3ab}(b) shows the output intensity distribution of a co-propagating (zero time-delay) low-power probe beam under the influence of the pump beam as a function of the pump input power (see also Eq.~\ref{eq:2}). For low pump power we observe probe beam intensities on the left and right of the pump waveguide, as expected for the excitation of extended, dispersed bulk states. With increasing pump input power, the power on the left of the pump beam decreases towards zero, as well as the power on the A sites to the right of the pump beam. This transition coincides with the pump forming the highly localized soliton, detuning the input site to form the `wall' between the left and right sides. At the highest pump input power that we can reach in our experiment (without damaging the sample), for which the wall is the most pronounced, we observe that the probe beam is populating B sites with more intensity than A sites, a hallmark of an SSH end state. 
Repeated measurements for different input waveguides show a similar behavior (see Supplementary Information Fig. S7). Note that the power threshold for the soliton bifurcation depends very sensitively on in-coupling efficiency, the slight refractive index differences among the waveguides, as well as their losses. Therefore, the induced SSH end state appears at different power thresholds for each measurement (see Supplementary Information for more details).

To quantitatively evaluate the end state formation, we analyze the intensity imbalance $(I_\mathrm{B}-I_\mathrm{A})/(I_\mathrm{B}+I_\mathrm{A})$ 
of the probe beam for the intensity $I_\mathrm{A}$ ($I_\mathrm{B}$) in the A (B) sites to the right of the probe beam for the measurements shown in Fig.~\ref{fig:3ab} (b), as shown in Fig.~\ref{fig:3c}. In the SSH model, chiral symmetry dictates that all bulk eigenstates have equal support on both sublattices, i.e., the imbalance for all bulk states is zero. By contrast, the topological end states at mid-gap have support on only one of the two sublattices. While a linear superposition of bulk states may be imbalanced, this will vanish with increasing propagation distance. A large imbalance therefore indicates the excitation of a topological end state. Note that although the soliton locally breaks chiral symmetry, chiral symmetry is still approximately present in the bulk - see Supplementary Information section IV. Fig.~\ref{fig:3c} shows a strong increase in the imbalance for increasing input pump power, signaling the creation of the topological end state due to the nonlinearly-induced hard wall. In contrast, we also probe the waveguide to the left of the pumped waveguide, site -1, i.e., its weakly coupled neighbor, which cannot host a topological end state (see Fig.~\ref{fig:0}(c)). The output intensity is plotted in Supplementary Figure S6 and the extracted imbalance (red line in Fig.~\ref{fig:3c}) remains around zero even for increasing pump power. This serves as an experimental control and confirms that the soliton induces an end state on its neighboring strongly-coupled site.

\section{Conclusion \label{Conclusion}}
In conclusion, we have shown theoretically and experimentally that the inversion symmetry-breaking nonlinear bifurcation that forms a soliton peaked on a single site in a nonlinear SSH lattice, spontaneously induces a topological end state on its neighboring strongly-coupled site. This work demonstrates the active formation and control topological states all-optically, anywhere in the bulk of the system. 




\begin{acknowledgments}
  C.J.\ thanks Debadarshini ``Jolly" Mishra and Sachin Vaidya for helpful discussions. We further acknowledge Nicholas Smith of Corning Inc. for providing high quality Eagle XG wafers, and are grateful to the Photonikzentrum Kaiserslautern for the use of their laser. C.J.\ acknowledges funding from the Alexander von Humboldt Foundation within the Feodor-Lynen Fellowship program. 
  M.C.R.\ and M.J.\ acknowledge the support of the U.S.\ Office of Naval Research under Grant No.\ N00014-23-1-2102, as well as the ONR Multidisciplinary University Research Initiative (MURI) under Grant No.\ N00014-20-1-2325. S.M. acknowledges support from Indian Institute of Science. 
\end{acknowledgments}

\bibliography{references}

\end{document}


\title{Supplementary Information: Soliton-induced topological end states in the bulk of an SSH lattice}
\author{Christina J{\" o}rg}
\affiliation{Department of Physics, The Pennsylvania State University, University Park, Pennsylvania 16802, USA}
\affiliation{Physics Department and Research Center OPTIMAS, University of Kaiserslautern-Landau, Kaiserslautern D-67663, Germany}
\author{Marius J{\" u}rgensen}
\affiliation{Department of Physics, The Pennsylvania State University, University Park, Pennsylvania 16802, USA}
\author{Sebabrata Mukherjee}
 \affiliation{Department of Physics, The Pennsylvania State University, University Park, Pennsylvania 16802, USA}
\affiliation{Department of Physics, Indian Institute of Science, Bangalore 560012, India}

\author{Mikael C. Rechtsman}
\affiliation{Department of Physics, The Pennsylvania State University, University Park, Pennsylvania 16802, USA}

\date{\today}

\maketitle

\section{Fabrication}
The waveguide arrays were fabricated in Corning Eagle XG borosilicate glass by femtosecond laser direct writing. We used a Menlo BlueCut fiber laser with a wavelength of \SI{1030}{nm} and repetition rate of \SI{500}{kHz}. The average laser power for the fabrication of the waveguides was around \SI{225}{mW} in the laser focus and the sample on the stage was translated at \SI{8}{mm/s} through the laser focus. We shaped the beam by the use of a slit with width of \SI{1.8}{mm} to control the elliptical cross-section of the waveguides (for more details see Supplementary Information section of reference \cite{SebaFloquetSoliton}).

\section{Measurement setup}
\begin{figure*}
    \centering
    \includegraphics[width=\linewidth]{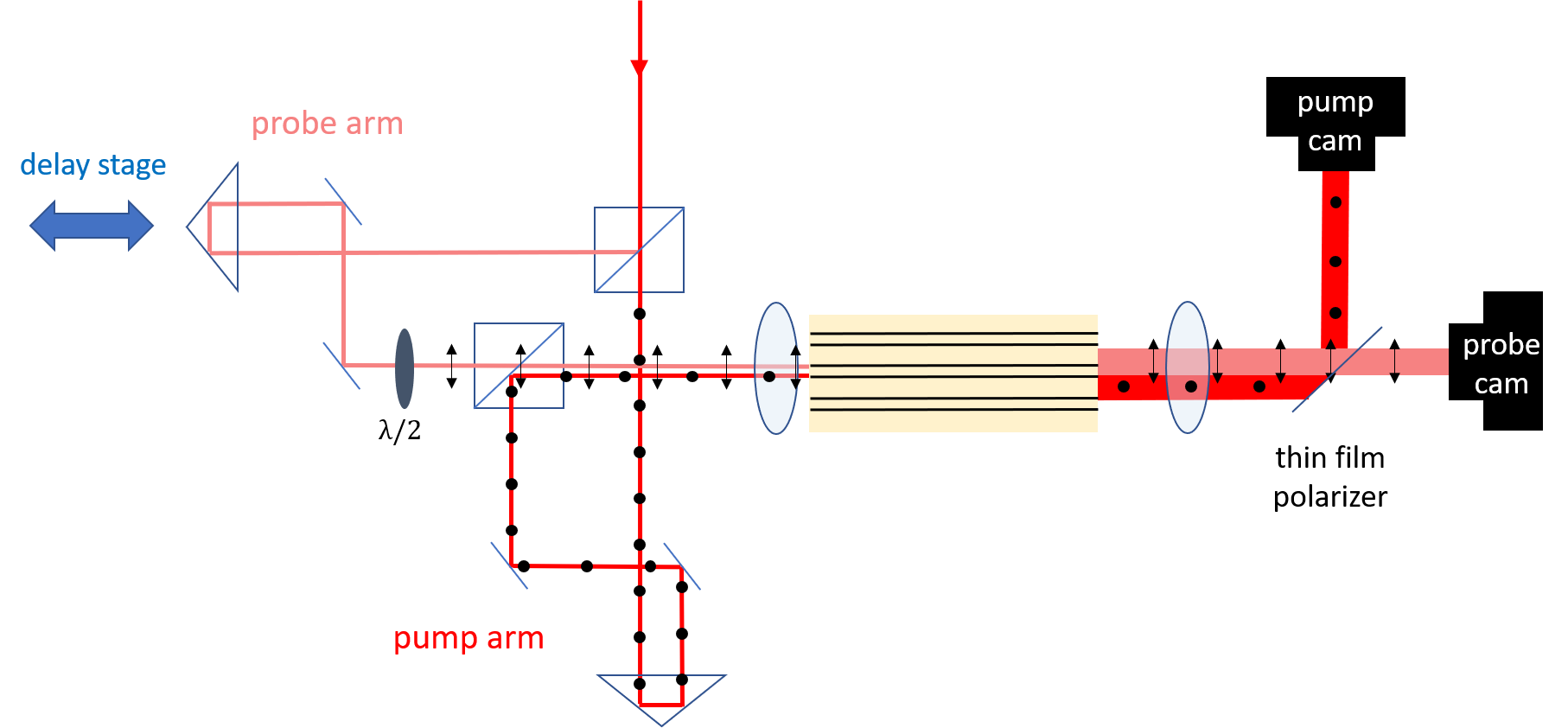}
    \caption{ \small{
        \textbf{Measurement setup.} A strong pump beam excites the soliton in the waveguide array, while a weak probe beam with orthogonal polarization probes the induced topological end state. 
      \label{fig:S1}}}
\end{figure*}

The measurement setup is depicted in Fig.~\ref{fig:S1} and shows a pump-probe setup in which pump and probe polarization can be tuned individually. A pulsed laser beam (Menlo BlueCut) is sent through a Glan Thompson polarizer which allows the power to be tuned, and is polarized horizontally by a wire-grid polarizer (not shown in Fig.~\ref{fig:S1}). The beam has a wavelength of \SI{1030}{nm} and pulse duration of approximately \SI{450}{fs} at a repetition rate of \SI{10}{kHz}. The beam is then split by a 90:10 non-polarizing beamsplitter cube into two paths, the probe path (light red), and the pump path (dark red). 
The polarization of the beam in the probe path is rotated by 90° by a half-waveplate. 
The beam in the probe path is retro-reflected by a prism mounted onto a delay stage, which allows to overlap the pulses of the pump and probe beam temporally by adjusting the probe beam's path length.
A second 50:50 non-polarizing beamsplitter cube combines the pump and probe beam again, before an aspheric lens ($f = \SI{18.40}{mm}$, Thorlabs C280TMD-B)  focuses the two beams into neighboring waveguides. The output intensity of the waveguides' output facet is imaged by a lens ($f = \SI{25.4}{mm}$, Thorlabs LB1761-B-ML), and we use a thin film polarizer under the Brewster angle to separate the pump and probe beam polarizations, such that they are imaged onto two separate CMOS-cameras (Thorlabs DCC1545M). Since the polarizer is not perfect, a small amount of pump beam polarization remains leaking through onto the probe camera, which is non-negligible given that the pump beam intensity is significantly greater than the probe. 
After blocking the residual pump polarization using an additional linear polarizer, the resulting pump polarization leaking through to the camera increases with pump power, but is smaller than 10\% up to an input power of $\SI{2}{mW}$, and reaches approximately 20\% for an input power of $\SI{3}{mW}$. The increase in leakage with pump power points to polarization changes due to a nonlinear process.

We observe spectral broadening of the pump pulses due to self-phase modulation in the waveguides to about $\SI{70}{nm}$ FWHM at \SI{2}{mW} input power after \SI{150}{mm} of propagation in the glass waveguides.

Due to the elliptical cross-section of the fabricated waveguides, the coupling constants between neighboring waveguides is a function of polarization, and thus different for pump and probe beam. The values for the coupling constants for the different polarizations are extracted from measurements on simple integer lattices and used in the simulations.


%

\section{Linear stability analysis}
We conduct linear stability analysis, following reference \cite{kevrekidis_discrete_2009}, to check the stability of the soliton solutions found by the self-consistency method. 
The solution $\Psi$ of the nonlinear Schrödinger equation with eigenvalue $E_0$
\begin{align}
    E_0 \Psi= H_\mathrm{lin}\Psi -g|\Psi|^2\Psi
    \label{eq:0}
\end{align}
is of the form $\Psi(t)=\Psi(x)\exp{(-\mathrm{i}E_0t)}$. 
Note that 

$\Psi=\begin{pmatrix}
   \Psi_1\\ 
   \Psi_2\\ 
   \vdots\\ 
   \Psi_N 
 \end{pmatrix}$
 is a vector containing the electric field amplitudes at the waveguide sites, and 

$g|\Psi|^2=\begin{pmatrix}
   g|\Psi_1|^2 &  &  & \\ 
   & g|\Psi_2|^2 &  & \\ 
   &  &  \ddots & \\ 
   &  &   & g|\Psi_N|^2 
 \end{pmatrix}$
 represents a diagonal matrix. 
We add a small perturbation to this solution
\begin{align}
    \Psi(t)=(\Psi(x)+\epsilon(v+\mathrm{i}w))\exp{(-iE_0t)}.
    \label{eq:1}
\end{align}
Inserting Eq.~\ref{eq:1} into Eq.~\ref{eq:0}, neglecting higher order terms in $\epsilon$ ($\mathcal{O}(\epsilon^2)$) and sorting for real and imaginary parts yields
\begin{align}
    \partial_t w
    =-(-E_0 -2g |\Psi|^2-g\Psi^2+H_\mathrm{lin})v=-L_+v
    \label{eq:2}
\end{align}
and
\begin{align}
    \partial_t v
    =(-E_0 -2g |\Psi|^2+g\Psi^2+H_\mathrm{lin})w=L_-w.
    \label{eq:3}
\end{align}
We separate time and space variables
\begin{align}
    v(x,t)=\Tilde{v}(x)\exp{(\lambda t)}\\
    w(x,t)=\Tilde{w}(x)\exp{(\lambda t)}
    \label{eq:6a}
\end{align}
and thus Eq.~\ref{eq:2} and \ref{eq:3} result in an eigenvalue problem
\begin{align}
    \lambda^2 \Tilde{v}=-L_- L_+ \Tilde{v}.
    \label{eq:4}
\end{align}
We see that eigenvalues of $-L_- L_+$ with $\lambda^2>0$ imply unstable soliton solutions $\Psi$.


\begin{figure*}
    \centering
    \includegraphics[width=\linewidth]{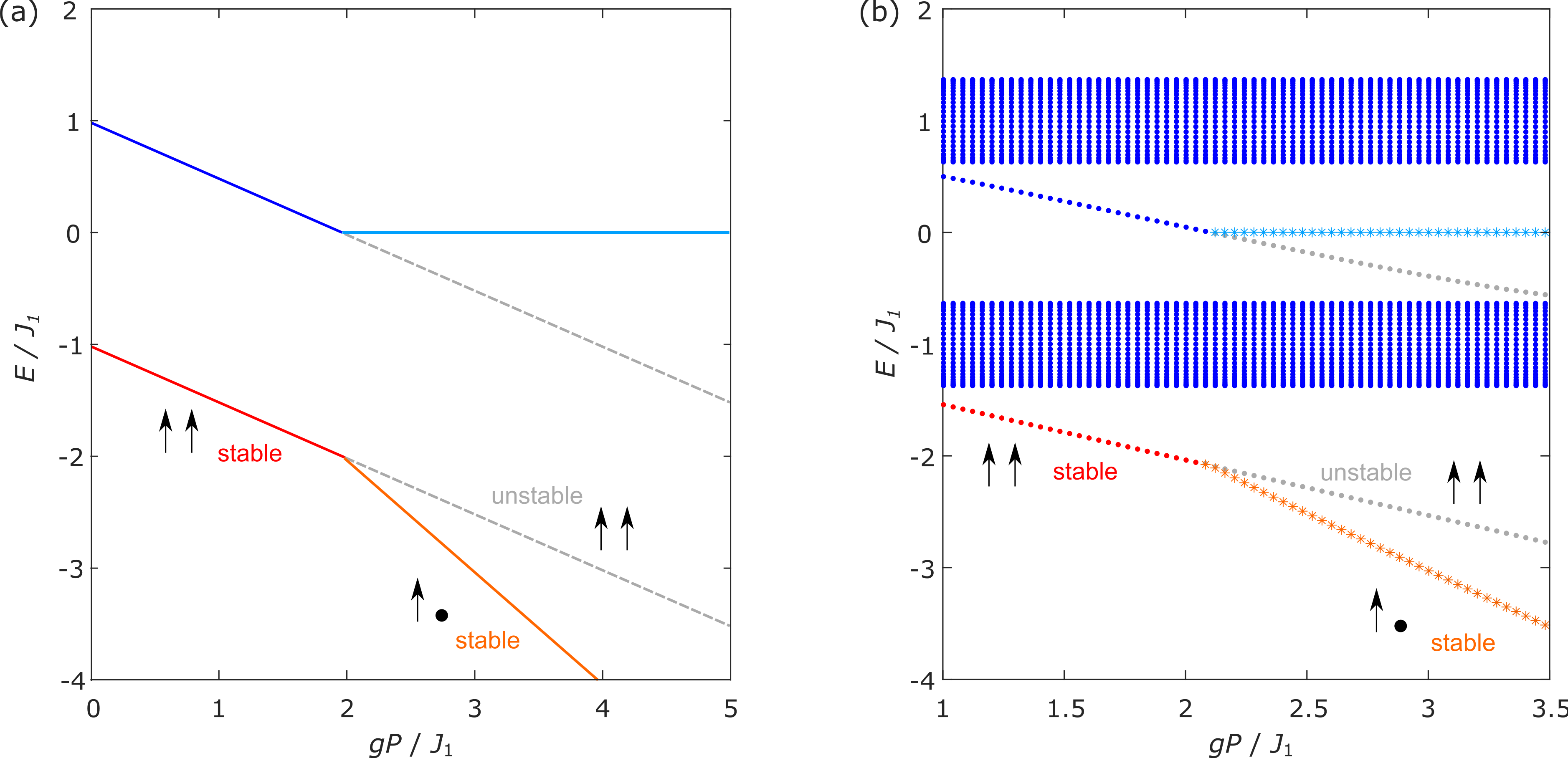}
    \caption{ \small{
        \textbf{Linear stability analysis.} Eigenvalue diagrams as a function of power for the case of a single dimer (a) and the SSH lattice with $J_2/J_1=0.37$ (b). Unstable soliton eigenvalues (with their corresponding in-gap states) are shown in gray. The symmetric soliton (denoted by $\upuparrows$ and red dots) is only stable up to power $gP/J_1<2$ (the bifurcation point). For larger power, this symmetric soliton becomes unstable and therefore the corresponding in-gap state ceases to exist (gray dots). Instead, a new stable soliton emerges with more support on one single site (denoted by \textuparrow $\bullet$ and orange stars), and its corresponding in-gap state (light blue stars) now is at $E=0$ for all $gP/J_1>2$.  
      \label{fig:S2}}}
\end{figure*}


Figures~\ref{fig:S2}(a),(b) show the eigenvalue spectra colored in terms of the linear stability eigenvalues of Eq.~\ref{eq:4}, where gray dots mean that the soliton is unstable ($\lambda^2>0$). Shown are the case of a single dimer (a), i.e. $J_2=0$, and the SSH lattice with $J_2/J_1=0.37$ (b).
For the single dimer in (a), we see that for low power a stable symmetric nonlinear eigenstate exists with equal phase and amplitude on both dimer sites (denoted by $\upuparrows$ and red dots): 
\begin{align}   
    \Psi^{\mathrm{pu}} = \sqrt{P/2} \left(
\begin{array}{c}
1\\
1\\
\end{array}
\right)
    \label{eq:S8}
\end{align}

\begin{figure}
    \centering
    \includegraphics[width=\linewidth]{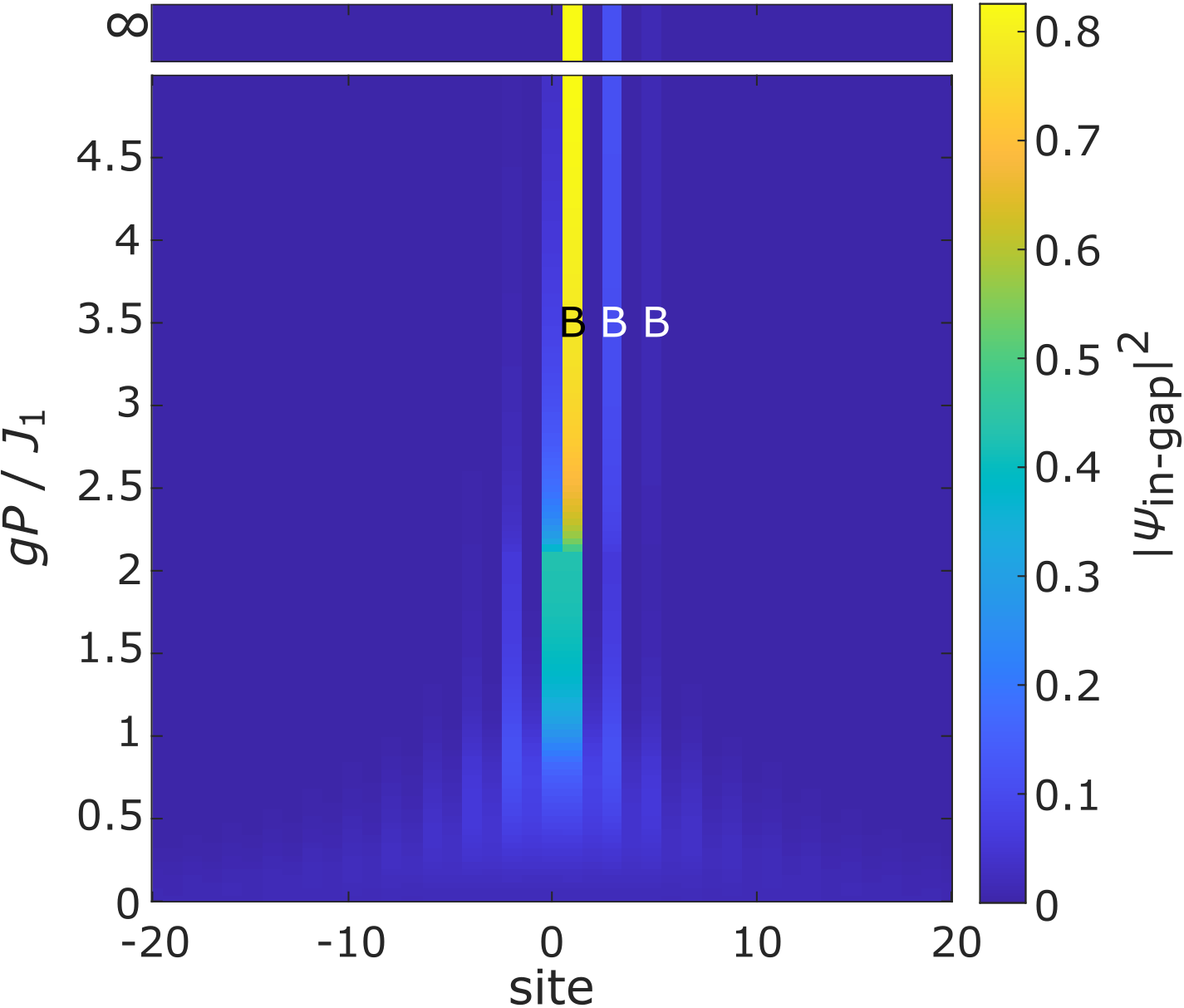}
    \caption{ \small{
        \textbf{Numerical simulations of the intensity of the in-gap state} $\psi_\mathrm{in-gap}$} as a function of power, as obtained via the self-consistency method. The letters in the plot indicate sublattice B.
      \label{fig:S3}}
\end{figure}

The eigenvalues of the probe states (the solutions of Eq. 2 in the main text) are plotted in blue, and correspond to the state
\begin{align}   
    \Psi^{\mathrm{pr}} = \sqrt{P/2} \left(
\begin{array}{c}
1\\
-1\\
\end{array}
\right).
    \label{eq:S9}
\end{align}
At the bifurcation point, the symmetric soliton becomes unstable for $P/J_1>2$ (lower gray dots) and therefore the corresponding probe state ceases to exist (upper grey dots). Instead, a new stable nonlinear eigenstate emerges with more support on one site than the other (denoted by \textuparrow $\bullet$ and orange stars) 
\begin{align}   
    \Psi^{\mathrm{pu}} = \sqrt{P/2} \left(
\begin{array}{c}
\sqrt{1+\sqrt{1-(2J/gP)^2}}\\
\sqrt{1-\sqrt{1-(2J/gP)^2}}\\
\end{array}
\right)
    \label{eq:S10}
\end{align}
\cite{eilbeck_discrete_1985}, and its corresponding probe state (light blue stars)
\begin{align}   
    \Psi^{\mathrm{pr}} = \sqrt{P/2} \left(
\begin{array}{c}
-\sqrt{1-\sqrt{1-(2J/gP)^2}}\\
\sqrt{1+\sqrt{1-(2J/gP)^2}}\\
\end{array}
\right)
    \label{eq:S11}
\end{align}
has eigenvalue $E=0$ for all $gP/J_1>2$, as one can easily check when inserting the states in Eq. 2 in the main text. 
A very similar bifurcation exists for the SSH lattice with $J_2/J_1=0.37$ in Fig.~\ref{fig:S2}(b): The symmetric soliton becomes unstable near $gP/J_1=2$, and the new stable soliton has more support on a single site \textuparrow $\bullet$ (see also Refs. \cite{PhysRevA.79.065801,Kanshu:12}). The corresponding in-gap state then also changes to have $E=0$.  This provides a direct link between the extreme dimer case in Fig.~\ref{fig:S2}(a) and the SSH case in Fig.~\ref{fig:S2}(b) - both occur due to an inversion-symmetry breaking bifurcation.  In the SSH case, this allows for the formation of an effective topological end next to the site with the soliton peak (see Fig.~\ref{fig:S3}).



\section{Approximate chiral symmetry}
\begin{figure}[t]
    \centering
    \includegraphics[width=\linewidth]{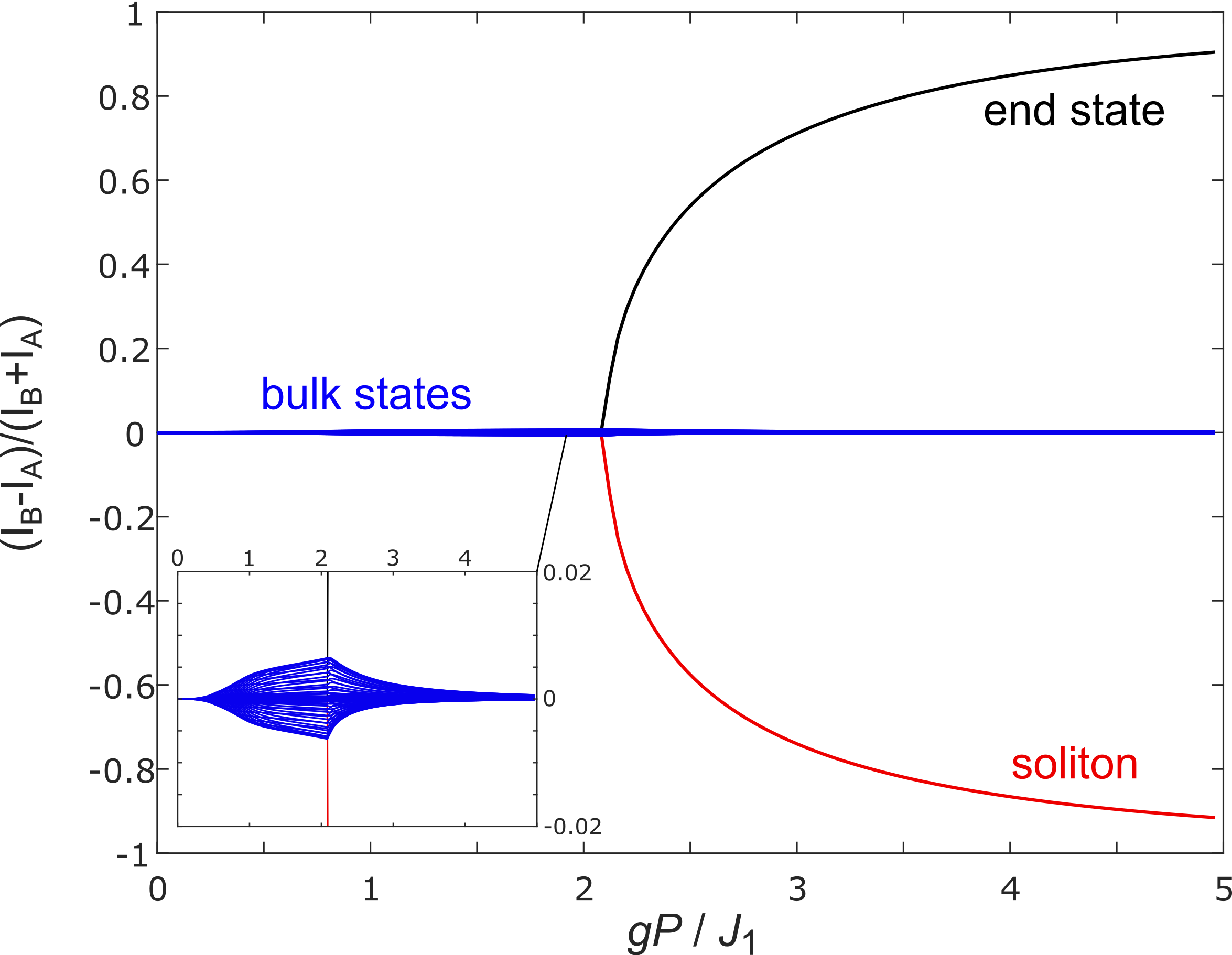}
    \caption{ \small{
        \textbf{Approximate chiral symmetry.}  
        Sublattice imbalance of the eigenstates of the nonlinear Hamiltonian (Eq. 1 in the main text) as a function of power. The imbalance for the soliton is plotted in red, the topological end state in black, and bulk states in blue. The imbalance for the bulk states is not exactly zero, as in the linear case, but very small (see also inset for same data, zoomed in on the bulk states), indicating that chiral symmetry is nearly preserved for the bulk states.
      \label{fig:S6}}}
\end{figure}

As stated in the main text, perfect chiral symmetry enforces zero sublattice imbalance for bulk eigenstates in the SSH lattice. In our experiments, the soliton breaks chiral symmetry via the potential it generates, but only locally. In order to quantify the degree of chiral symmetry breaking, we calculate the sublattice imbalance of the eigenstates of the nonlinear Hamiltonian (Eq. 1 in the main text) as a function of power, see  Fig.~\ref{fig:S6}. The imbalance for the soliton (red in Fig. 2 of the main text) and topological end state are plotted in red and black respectively, while the blue lines mark the imbalance for bulk eigenstates. For increasing power the imbalance for the bulk states deviates from zero (zero means perfect chiral symmetry), but its deviation is small (below 1\%, see inset of Fig. S4, zoomed in on the bulk states) for all soliton powers, indicating that chiral symmetry is nearly preserved for the bulk states.

As stated in the main text, for single-site input we excite a superposition of multiple bulk and possible end states. Therefore, the imbalance for a topologically trivial lattice does not strictly vanish, but decreases as a function of the propagation distance $z$.

\begin{figure}
    \centering
    \includegraphics[width=\linewidth]{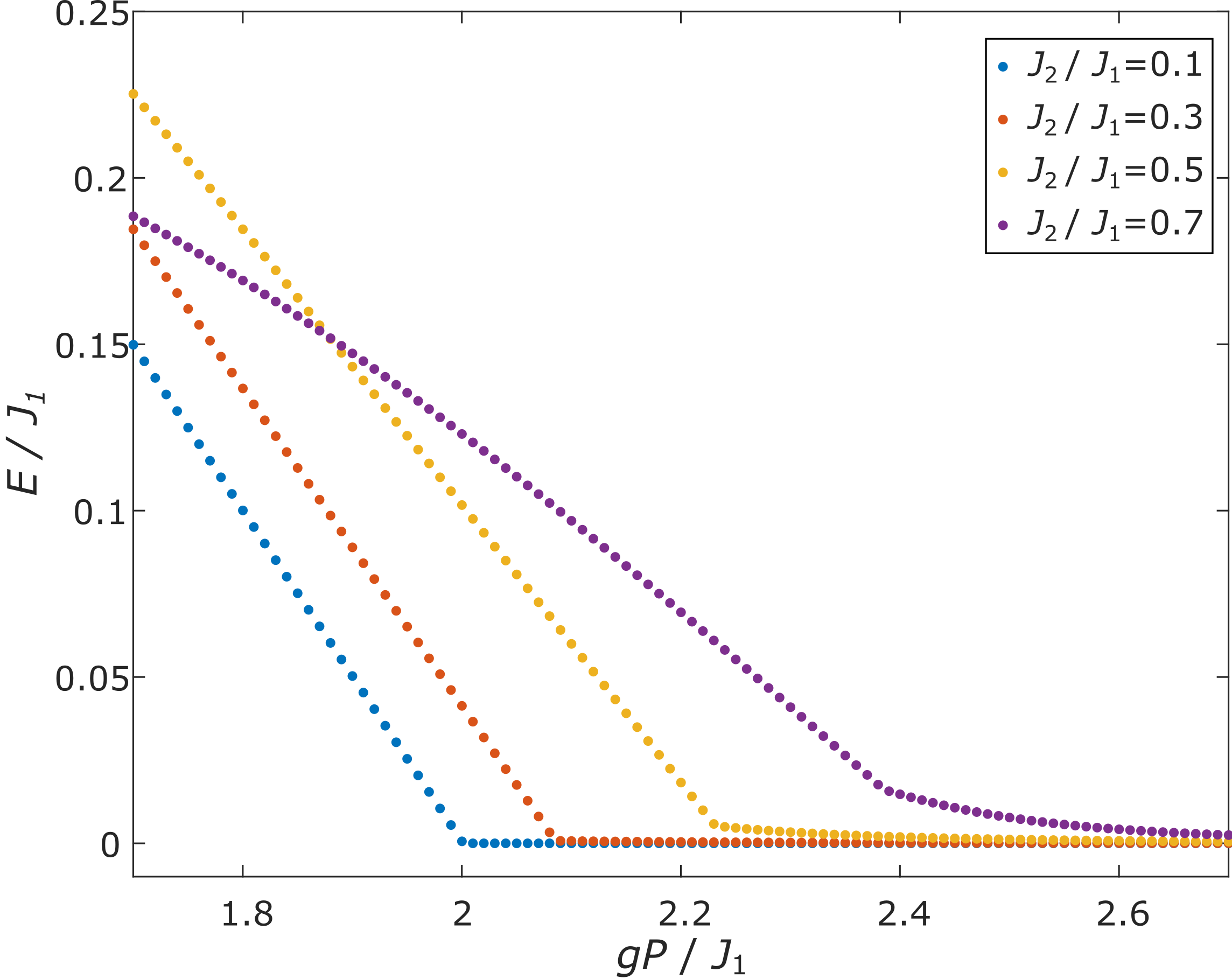}
    \caption{ \small{
        \textbf{Bifurcation for increasing $J_2$,} zoom-in on the in-gap eigenvalues.  
        For increasing $J_2$ the bifurcation point of the soliton as well as the transition point of the in-gap state shifts to higher power and happens more gradually. 
      \label{fig:S4}}}
\end{figure}

\section{Transition for increasing $J_2$}
We examine the bifurcation transition as we deviate from a perfectly dimerized lattice ($J_2=0$) by increasing the second coupling $J_2$.
Figure~\ref{fig:S4} shows the transition for multiple values of $J_2$. For increasing $J_2$ the bifurcation point of the soliton as well as the transition point of the in-gap state shifts to slightly higher power and the in-gap state's eigenvalue goes to zero more gradually.

\section{Additional measurements}
\begin{figure*}
\centering
    \includegraphics[width=\linewidth]{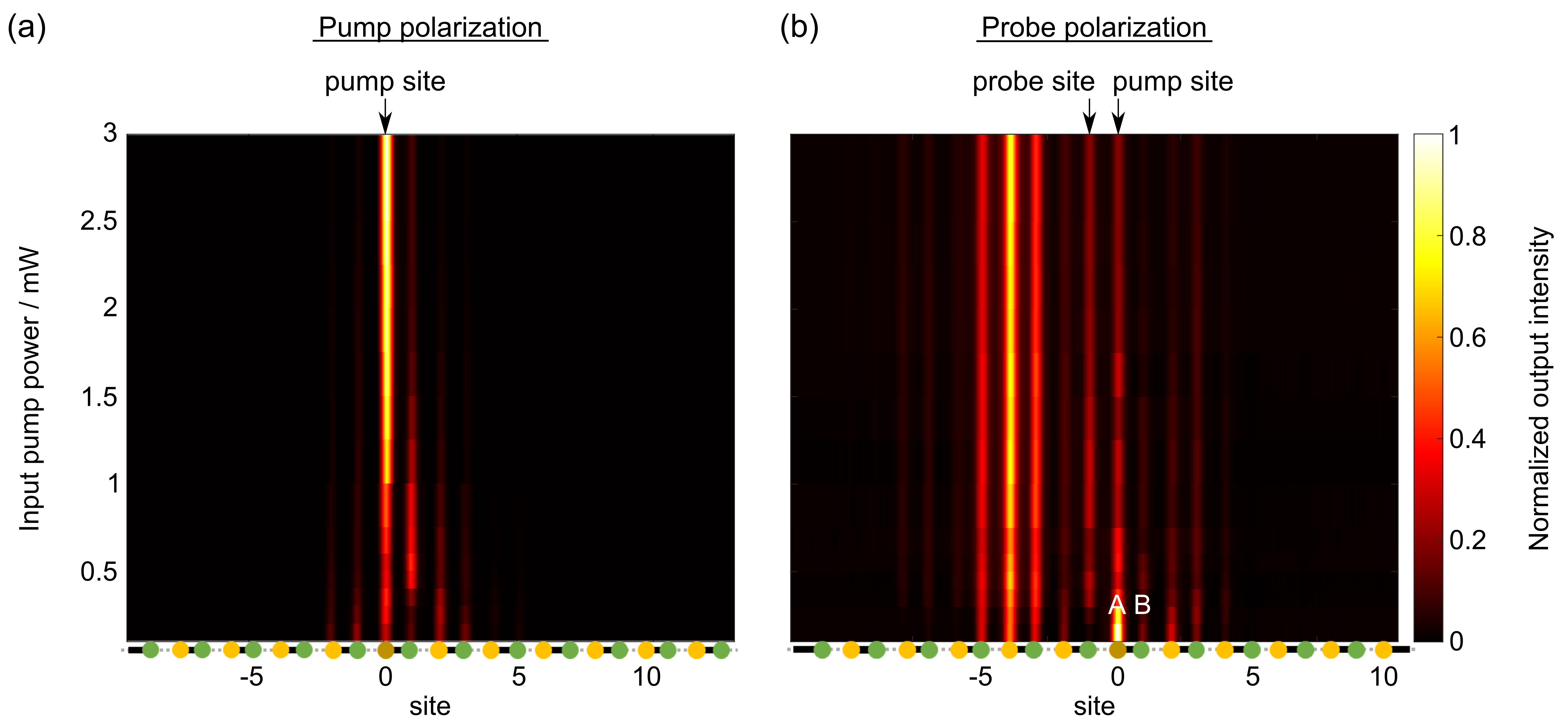}
    \caption{ \small{
        \textbf{Probing the trivial ending} by probing waveguide -1 to the left of the induced soliton. (a) Pump polarization, (b) probe polarization. 
      \label{fig:S8}}}
\end{figure*}
\begin{figure*}
\centering
    \includegraphics[width=\linewidth]{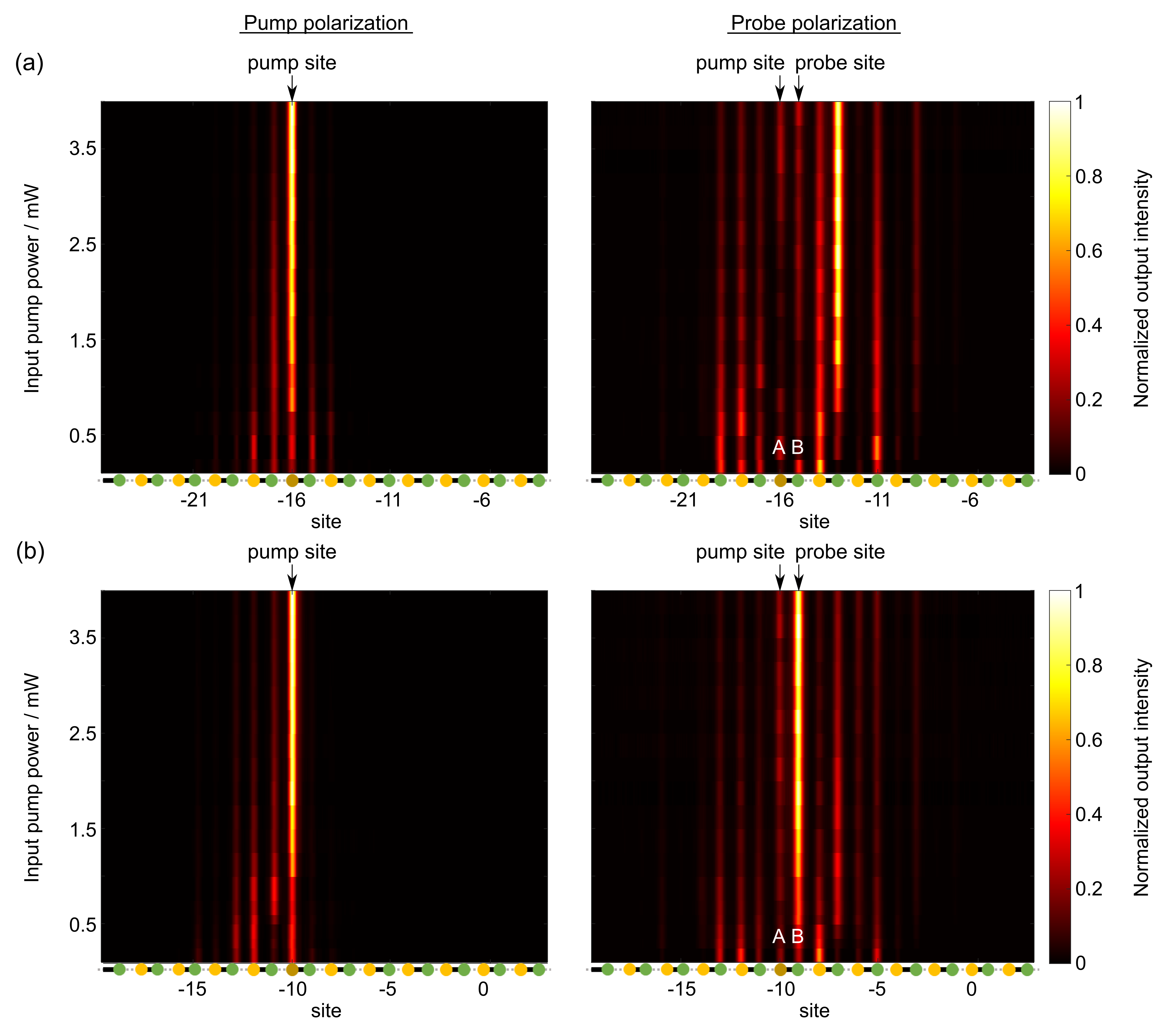}
    \caption{ \small{
        \textbf{Additional measurements at different positions in the lattice.} Left: Pump polarization, right: probe polarization. (a) and (b) are two different measurements at different positions in the lattice, conducted the same way as for Fig. 3 of the main text.
      \label{fig:S9}}}
\end{figure*}
\begin{figure*}
    \centering
    \includegraphics[width=\linewidth]{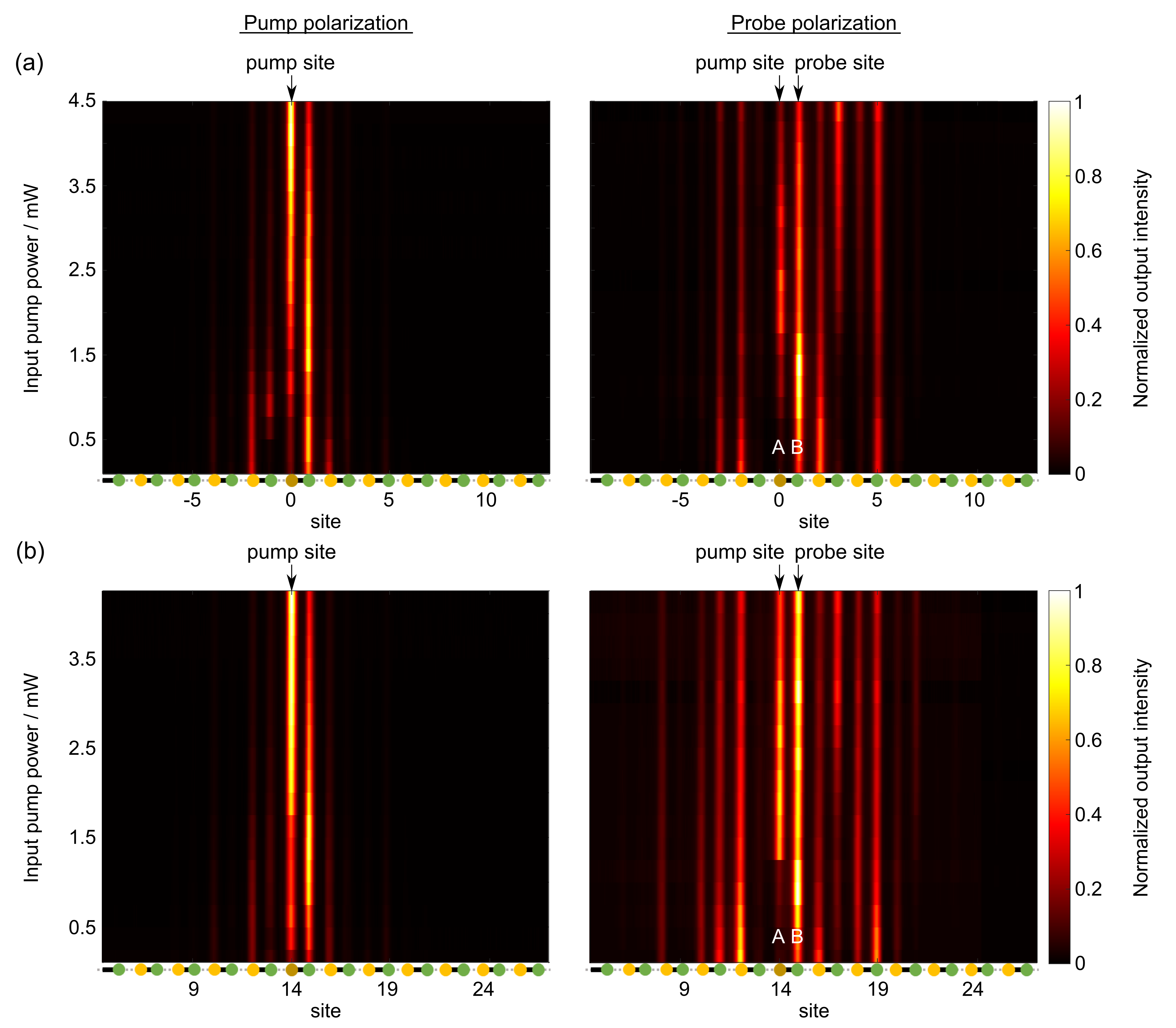}
    \caption{ \small{
        \textbf{Measurements when sample is flipped by $180^{\circ}$ around the z-axis}. Due to fabrication, one of the sublattices has a slightly higher refractive index than the other, which results in a higher power for the bifurcation threshold when the site with lower refractive index is pumped, as done here (compare Fig.~\ref{fig:S9}). Left: Pump polarization, right: probe polarization. (a) and (b) are two different measurements at different positions in the lattice.
      \label{fig:S7}}}
\end{figure*}

Fig.~\ref{fig:S8} shows the output intensity when probing the waveguide to the left of the pumped waveguide, site -1 (i.e., its weakly coupled neighbor). Since it cannot host a topological end state, we do not observe an intensity imbalance across A and B sites.

In Fig.~\ref{fig:S9} we show additional measurements at different positions in the lattice. In all cases, for high input power the intensity in the B-sites to the right of the probe waveguide is noticeably higher than the intensity in the A-sites. 

\begin{figure*}
    \centering
    \includegraphics[width=\linewidth]{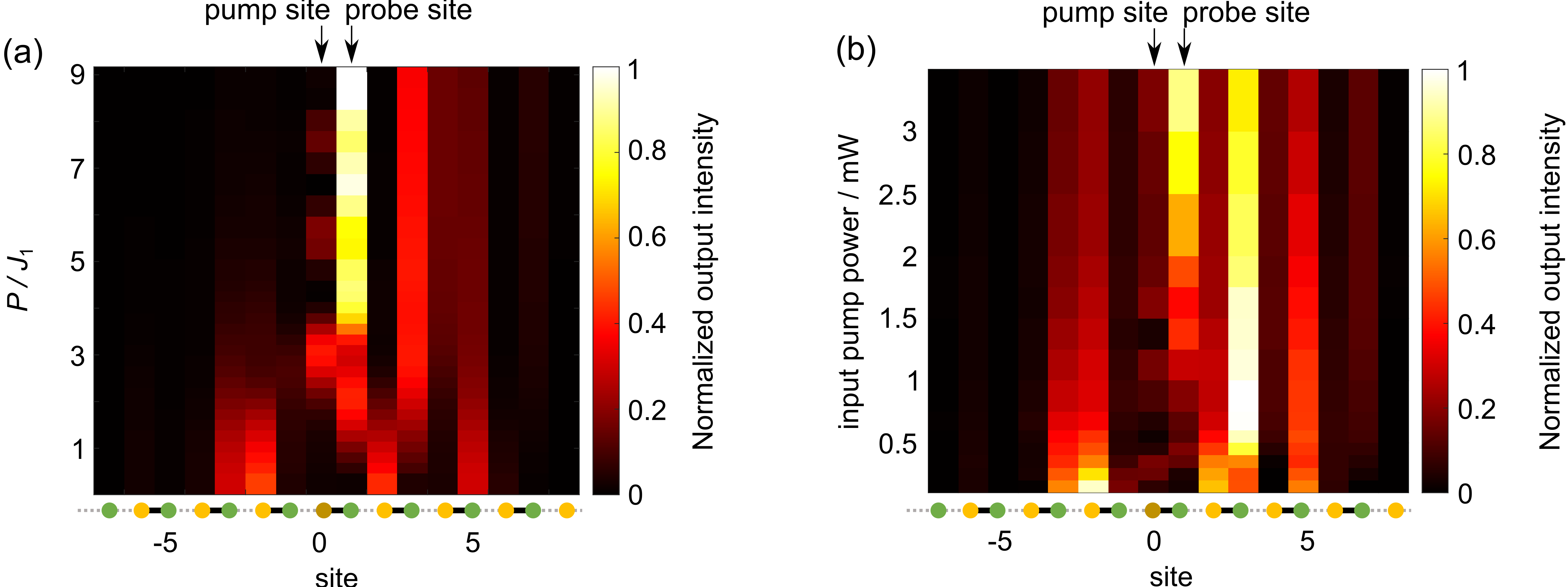}
    \caption{ \small{
        \textbf{Comparison of Runge-Kutta simulations (a) with the experimental data (b).} Shown is the output intensity in the probe polarization. The Runge-Kutta simulations contain an on-site potential of $+0.8J_1$ ($-0.8J_1$) on the A (B) sites. In (b) only the peak intensity per waveguide site is plotted.
      \label{fig:S10}}}
\end{figure*}

\begin{figure*}
    \centering
    \includegraphics[width=\linewidth]{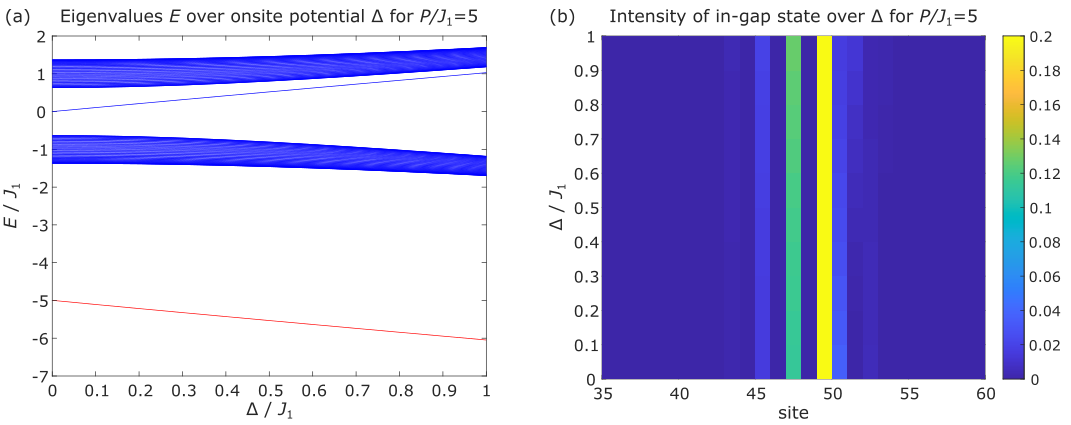}
    \caption{ \small{
        \textbf{Influence of the onsite potential difference:} (a) The eigenvalue of the in-gap state increase from zero (mid-gap) for increasing onsite potential $\Delta/J_1$, while the intensity profile of the in-gap state stays almost the same (b). Both plotted for a power value of $gP/J_1=5$, well above the bifurcation point.
      \label{fig:S11}}}
\end{figure*}

When pumping site B (instead of A) and probing site A (instead of B), we observe that the power threshold for the bifurcation noticeably increases (Fig. S8). Experimentally, we achieve this by keeping our beams identical, but flipping the sample by 180° around the z-axis. This increase of the bifurcation threshold points to a slight on-site potential difference between A and B waveguides due to fabrication. In the Runge-Kutta simulations (Fig.~\ref{fig:S10}), we therefore included an onsite potential of $\Delta=+0.8J_1$ ($-0.8J_1$) on the A (B) sites. Fig.~\ref{fig:S11} shows that the onsite potential has negligible effect on the calculated intensity profile of the in-gap state.
One should note that additional effects, such as losses in the waveguides and the fact that part of the input light consists of lower power tails of the input pulse, are not captured in simulations. 

\bibliography{supplement_references2}